\documentclass[%
 reprint,
 superscriptaddress,
 amsmath,amssymb,
 aps,
prb,
floatfix,
longbibliography 
]{revtex4-2}

\usepackage[pdftex]{graphicx} \graphicspath{{}}
\usepackage{float}
\usepackage{ulem}
\usepackage{dcolumn}
\usepackage{bm}
\usepackage[pdftex,colorlinks=true]{hyperref}
\hypersetup{
  colorlinks=true,
  linkcolor=blue,
  urlcolor=cyan,
}
\usepackage[section]{placeins}

\newcommand{\eqn}[1]{\begin{equation} #1 \end{equation}}
\newcommand{\eqa}[1]{\begin{align} #1 \end{align}}
\def\be{\begin{equation}}
\def\ee{\end{equation}}

\usepackage{color}

\newcommand{\nn}{\nonumber}

\newcommand{\mH}{\mathcal{H}}
\newcommand{\mD}{\mathcal{D}}

\newcommand{\avg}[1]{\left\langle #1 \right\rangle}

\newcommand{\bS}{\boldsymbol{S}}

\newcommand{\mC}{\mathcal{C}}
\newcommand{\mA}{\mathcal{A}}

\newcommand{\bJ}{\boldsymbol{J}}

\newcommand{\bj}{\boldsymbol{j}}

\DeclareMathOperator{\eff}{eff}
\DeclareMathOperator{\KPZ}{KPZ}

\newcommand{\sectionn}[1]{\textit{#1---}}

\begin{document}

\preprint{APS/123-QED}

\title{Parametrically long lifetime of superdiffusion in non-integrable spin chains}%

\author{Adam J. McRoberts}
\affiliation{International Centre for Theoretical Physics, Strada Costiera 11, 34151, Trieste, Italy}
\affiliation{Max Planck Institute for the Physics of Complex Systems, N\"{o}thnitzer Str. 38, 01187 Dresden, Germany}

\author{Roderich Moessner}
\affiliation{Max Planck Institute for the Physics of Complex Systems, N\"{o}thnitzer Str. 38, 01187 Dresden, Germany}

 
\date{\today}

\begin{abstract}
\noindent
Superdiffusion is surprisingly easily observed even in systems without the integrability underpinning this phenomenon. Indeed, 
the classical Heisenberg chain -- one of the simplest many-body systems, and firmly believed to be non-integrable -- evinces a long-lived regime of anomalous, superdiffusive spin dynamics at finite temperature. Similarly, superdiffusion persists for long timescales, even at high temperature, for small perturbations around a related integrable model. Eventually, however, ordinary diffusion is believed to be asymptotically restored. We examine the timescales governing the lifetime of the superdiffusive regime, and argue that it diverges algebraically fast -- both in deviation from the integrable limit, and at low temperature, where we find $t^* \sim T^{-\zeta}$ with an exponent possibly as large as $\zeta = 8$. This can render the crossover to ordinary diffusion practically inaccessible.
\end{abstract}

\maketitle

\sectionn{Introduction}
The hugely successful general theory of hydrodynamics posits that the long-time dynamics of generic many-body systems is universally determined by their conservation laws and symmetries \cite{Chaikin_Lubensky}. In particular, conserved quantities generally exhibit diffusion. 

The remarkable discovery of superdiffusion in {\it integrable} spin chains with non-abelian symmetries \cite{das2019kardar, Krajnik2020, bulchandani2020KPZ, de2021stability, Ilievski_2021, gopalakrishnan2023superdiffusion, gopalakrishnan2023anomalous, weiner2020high, de2019anomalous, ilievski2016string, ilievski2018superdiffusion, gopalakrishnan2019kinetic, ljubotina2019KardarParisiZhang, dupont2020universal, dupont2021spatiotemporal, ye2022universal} is not at variance with this statement. As integrability is non-generic, the expectation is that its breaking, no matter how weak, will restore conventional diffusive behaviour \cite{friedman2020diffusive,durnin2021nonequilibrium, bertini2015prethermalization, bastianello2021hydrodynamics, vznidarivc2020weak, lopez2021hydrodynamics, glorioso2020hydrodynamics, lopez2023integrability, rabson2004crossover, lopez2022integrability}. 

The timescale of the crossover from integrabilty to genericity is, then, from the universalist perspective, a question of non-universal detail. But from a practical perspective of actually observable physical phenomena, the shoe can be on the other foot: even for a system as simple as the classical Heisenberg chain, it has been found that the crossover may in practice occur on a scale unlikely to be reached experimentally on any presently existing platform \cite{mcroberts2022anomalous, roy2023robustness, rosenberg2023dynamics, scheie2021detection, keenan2023evidence, morningstar2022anomalous, wei2022quantum, joshi2022observing, zu2021emergent}. 

To make the two worldviews meet, the question which must be addressed is how the crossover timescale $t^*$ depends on various parameters, such as the temperature $T$, or the degree of integrability breaking $\delta$ in the Hamiltonian. 

This letter is devoted to such an investigation. We use the aforementioned classical Heisenberg chain as our model system for the following reasons: first, it is already established as one of the paradigmatic model systems in statistical physics, and its simplicity allows for relatively tractable numerical simulations; second, its near-integrable behaviour and anomalous dynamics have been subject to decades of debate~\cite{gerling1989comment, gerling1990time, bohm1993comment, srivastava1994spin, bagchi2013spin, muller1988anomalous, de1992breakdown, de1993alcantara, de2020universality, mcroberts2022anomalous}; and, third, we can smoothly interpolate between the Heisenberg chain and the integrable Ishimori chain, Eq.~\eqref{eq:H_Ishimori}, providing a reference integrable model on which the discussion can be based. 

We find that the growth of the crossover time can be parametrically very fast -- in particular, that it scales like $t^* \sim \delta^{-3}$ with the strength of the integrability-breaking perturbations; and, in the classical Heisenberg chain, its scaling with temperature is consistent with $t^* \sim T^{-8}$ -- though we caution the reader that, in a given temperature window, and depending on the strength and nature of the integrability-breaking, there may be various competing processes leading to a host of possible crossover phenomena -- non-universal physics, alas, can depend on details.

This account is structured as follows. We begin by considering the dynamics of the integrable Ishimori chain; this will lay the foundations for our subsequent analysis of the dynamics of its non-integrable deformations, including the nearest-neighbour Heisenberg chain. We will thus first consider the spin dynamics at high temperature (i.e., with random initial configurations) near the integrable point, and present a picture describing the crossover to normal diffusion in terms of the decay of solitons -- this predicts that the superdiffusive regime becomes parametrically long-lived in $\delta$ near the integrable limit, which we verify numerically. We then turn to the finite temperature spin dynamics in the classical Heisenberg chain and show that the superdiffusive regime in this model also becomes parametrically long-lived at low temperature, diverging with a large power law in $1/T$.

\sectionn{Superdiffusion in the integrable Ishimori chain}
Our integrable reference system is the Ishimori chain~\cite{ishimori1982integrable},
\eqn{
\mH = -2J \sum_i \log \left(\frac{1 + \bS_i \cdot \bS_{i+1}}{2} \right),
\label{eq:H_Ishimori}
}
where $\bS_i \in S^2$ are classical spins, with ferromagnetic interaction $J > 0$; we set $J = 1$ in the following, which implicitly defines all units. The classical equations of motion follow from the fundamental spin Poisson brackets, ${\{S^{\mu}_i, S^{\nu}_j\} = \epsilon^{\mu\nu\lambda}\delta_{ij}S^{\lambda}_j}$.

As in the integrable quantum $S = 1/2$ Heisenberg chain~\cite{ljubotina2019KardarParisiZhang}, the two-point spin correlator of the Ishimori chain, ${\mathcal{C}^S(j, t) := \avg{\bS_j(t)\cdot\bS_0(0)}}$, follows the KPZ scaling form~\cite{das2019kardar, ertacs1992dynamic, ertacs1993dynamic, prahofer2004exact}. That is,
\eqn{
\mathcal{C}^S(x, t) = \frac{\chi}{(\lambda_{\KPZ}t)^{2/3}} \mathcal{F}_{\KPZ}\left(\frac{x}{(\lambda_{\KPZ}t)^{2/3}}\right),
\label{eq:KPZ_corr}
}
where $\chi(T) = \sum_j \avg{\bS_j\cdot\bS_0} = 1 + 2/T$ is the static spin susceptibility, and $\lambda_{\KPZ}$ is the KPZ coupling. Whilst recent theoretical~\cite{de2023nonlinear} and experimental~\cite{rosenberg2023dynamics} work has shown that higher correlation functions may not follow the KPZ expectations (which would place the quantum Heisenberg and classical Ishimori chains outside the KPZ universality class), for our purpose of deriving superdiffusive timescales, it suffices to consider only two-point correlations, and, for lack of a better name, we  refer to this as ``KPZ-like'' superdiffusion.

\begin{figure}
    \centering
    \includegraphics{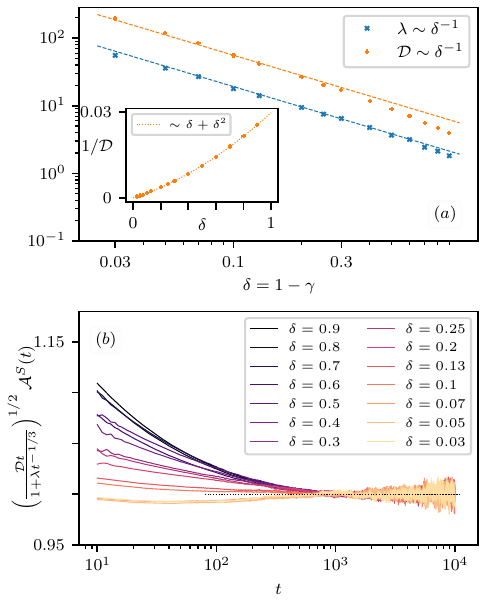}
    \caption{Long-lived superdiffusion for the non-integrable spin chains \eqref{eq:H_gamma}, as a function of integrability-breaking $\delta$, at $T = \infty$. In all cases  we use system size $L = 8192$. (a) $\delta$-dependence of the fitting parameters $\lambda$ and $\mathcal{D}$, with both showing the  $\sim 1/\delta$ scaling predicted in Eq.~\eqref{eq:crossover} near the integrable point $\delta = 0$ (dashed lines are fits to $1/\delta$). (Inset) Diffusion constant, evidencing a $1/\mD \sim \delta^2$ contribution as arises  
     {from large solitons scattering off small excitations (see text)} for moderate values of $\delta$. (b) The spin autocorrelator for various $\delta$, rescaled according to the crossover form \eqref{eq:crossover}, showing that the spin correlations are well-described by this form for $t \gtrsim 10^2 J^{-1}$, particularly near the integrable limit.
    }
    \label{fig:gamma_superdiffusion}
\end{figure}

\sectionn{Superdiffusion from large solitons}
To derive the superdiffusive timescales of the non-integrable chains we, in essence, extend  arguments found in Ref.~\cite{de2021stability}, which we  reproduce here for completeness. It is argued that the spin superdiffusion is attributable to large quasiparticles: magnon-strings in the $S=1/2$ Heisenberg chain \cite{de2021stability, de2020superdiffusion}; solitons in the Ishimori chain \cite{ishimori1982integrable}.

These `bound strings' of $s$ magnons (or, equivalently, solitons of width $\sim s$; the distinction vanishes at large $s$~\cite{de2020superdiffusion}) have dressed energy ${\varepsilon_s^{\mathrm{dr}} \sim s^{-1}}$, magnetisation ${m^{\mathrm{dr}}_s \sim s}$, and speed ${v_s^{\mathrm{dr}} \sim s^{-1}}$. Their thermal density scales as $\rho_s \sim s^{-3}$. Whilst all of these quasiparticles move ballistically (giving rise to ballistic energy transport), ballistic spin transport is arrested by the fact that when one $s$-quasiparticle collides with a larger quasiparticle $s' > s$, its magnetisation rotates to the local vacuum of the larger quasiparticle~\cite{mcroberts2022long}. Over many such collisions and rotations, the quasiparticle transports, on average, no net magnetisation; as argued in Ref.~\cite{de2021stability}, this may be modelled by exponential decay over a screening lifetime $\tau_s \sim (\rho_{s'>s} v_s)^{-1} \sim s^3$, where $\rho_{s'>s} = \int_s^{\infty} ds' \rho_{s'} \sim s^{-2}$ is the total density of quasiparticles with $s' > s$.

The local spin current $\bj^S_i$ is implicitly defined by the continuity equation, $\dot{\bS}_i = -\bj^S_{i+1} + \bj^S_i$. Defining the total spin current ${\bJ^S(t) = \sum_i \bj^S_i(t)}$, 
the above considerations imply that its correlator with the local current 
scales as
\eqa{
\left\langle\bJ^S(t)\cdot \bj^S_0(0)\right\rangle &\sim \int_0^{\infty} ds\; (m^{\mathrm{dr}}_s v_s^{\mathrm{dr}})^2\rho_s\, e^{-t/\tau_s} \nn \\
&\!\!\!\!\!\!\!\!\!\!\!\!\!\!\!\!\sim \kappa \int_0^{\infty} ds\; s^{-3} e^{-\eta t/ s^3} \sim \kappa (\eta \, t)^{-2/3}\ ,
\label{eq:KPZ_current}
}
for some constants $\kappa$ and $\eta$. The effective diffusion coefficient thus has the superdiffusive time-dependence
\eqn{
D(t) = \int_0^t dt' \avg{\bJ^S(t')\cdot \bj^S_0(0)} \sim \kappa\eta^{-2/3}t^{1/3}\ .
\label{eq:KPZ_diffusion_coefficient}
} 
This implies a decay of the spin autocorrelator $\mA^S(t) := \mC^S(0, t)$ with the desired KPZ exponent $\alpha = 2/3$,
\eqn{
\mA^S(t) = \frac{\chi}{\sqrt{D(t)t}} \sim t^{-2/3}.
\label{eq:spin_autocorr}
}

\sectionn{Superdiffusion near an integrable point}
We now consider how this picture changes when the Hamiltonian \eqref{eq:H_Ishimori} is modified with integrability-breaking terms, and ordinary diffusion is asymptotically restored. Concretely, we consider a model which interpolates between the integrable Ishimori chain \eqref{eq:H_Ishimori} and the non-integrable Heisenberg chain \eqref{eq:H_Heisenberg},
\eqn{
\mH = - 2J \gamma^{-1} \sum_i \log \left(1 + \gamma \frac{\bS_i \cdot \bS_{i+1} - 1}{2} \right),
\label{eq:H_gamma}
}
which recovers the Ishimori chain at $\gamma = 1$, and the Heisenberg chain in the limit $\gamma \rightarrow 0$, preserving $SO(3)$ symmetry throughout; $\delta = 1 - \gamma$, then, parametrises the degree of integrability-breaking. For now, we  work in the high-temperature limit to  separate this from the low-temperature superdiffusive timescales considered later.

Whilst we do expect  the dynamics of Eq.~\eqref{eq:H_gamma} to eventually cross over to ordinary spin diffusion, the large solitons responsible for the KPZ-like superdiffusion in the Ishimori chain are adiabatically connected to large solitons in the non-integrable chains \eqref{eq:H_gamma}~\cite{mcroberts2022long}, and thus source a long-lived regime of superdiffusive spin correlations. Related non-integrable models have also been shown to exhibit long-lived superdiffusion~\cite{roy2023robustness,roy2023nonequilibrium}.

It is now necessary to determine which process is most efficient at causing the large solitons to decay, and terminating superdiffusion. One might first consider processes which correspond to a single large soliton decaying as it propagates through a background of smaller excitations and spin waves -- which we can treat as an effective noisy perturbation to the dynamics. However, Ref.~\cite{de2021stability} considered noisy perturbations in the context of adding explicit time-dependent noise terms to the Hamiltonian, and observed that such processes are suppressed by Goldstone physics. This bounds the decay rate above by $\Gamma \sim 1/s^2$, and, consequently, these processes are not sufficient to drive a crossover to true ordinary diffusion -- they do suppress the superdiffusion in the sense that the exponent $\alpha \to 1/2$, but the diffusion coefficient diverges logarithimically at late times, $D(t) \sim \log(t)$. We refer to this as logarithmically-enhanced diffusion.

The next candidate is soliton decay via two-soliton scattering, with the decay rate set by the following considerations. Scattering is particularly destructive if the solitons significantly overlap during the collision, that is, if they are parametrically of the same size $\sim s$; if one of the solitons is much smaller than the other, the smaller will locally see a ferromagnetic ground state throughout the collision and tends to pass through largely unscathed. The density of ``similarly sized'' solitons scales as, say, $\int_{s/2}^{2s} s^{-3} ds \sim s^{-2}$~\footnote{The numerical factors $1/2$ and $2$ are, of course, somewhat arbitrary -- but since we are using this only to determine the scaling we could replace them with any others without affecting the result.}, and the rate at which such collisions occur then scales as $s^{-2}v_s \sim s^{-3}$. Now, in the quantum formulation of the problem, the matrix elements of the decay processes scale with $\delta$, and with the size of the overlap $\sim s$, so the decay probability in a given collision scales as $\delta^2 s^2$~\footnote{We are invoking a Fermi's golden rule-type argument for a classical spin chain---but we expect it to work, at least, in a long-wavelength setting where the quantum and classical models should be largely equivalent.}. Combining this with the collision rate, we conjecture that the soliton decay rate scales as $\Gamma \sim \delta^2/s$.
Following Ref.~\cite{de2021stability}, this implies that we should modify Eq.~\eqref{eq:KPZ_current} to
\eqn{
\avg{\bJ^S(t)\cdot \bj^S_0(0)} \sim \kappa \int_0^{\infty} ds\; s^{-3} e^{-\eta t/ s^3} e^{-\mu \delta^2 t/s}, 
\label{eq:crossover}
}
for some constant $\mu$.
The diffusion coefficient and the autocorrelator are then well-approximated by the following general crossover form, which matches the results of the above integral at early and late times,
\eqn{
D(t) \sim \frac{\mathcal{D}(\delta)}{1 + \lambda(\delta)\,t^{-1/3}}, \;\;\; \mA(t) = \frac{1}{\sqrt{D(t) t}}\,
\label{eq:diffusion_coefficient_crossover}
}
where
\eqn{
\mD(\delta) \sim \frac{\kappa}{\eta^{1/2}\mu^{1/2} \delta}, \;\;\; \lambda(\delta) \sim \frac{\eta^{1/6}}{\mu^{1/2}\delta}
\label{eq:crossover_parameters}
}
(note that in the integrable limit $\delta \to 0$, the diffusion coefficient in Eq.~\eqref{eq:diffusion_coefficient_crossover} with the parameters from Eq.~\eqref{eq:crossover_parameters} reduces to Eq.~\eqref{eq:KPZ_diffusion_coefficient}). We perform numerical simulations to test these predictions, and present the results in Fig.~\ref{fig:gamma_superdiffusion} (See App.~A of Ref.~\cite{mcroberts2024anomalous} for simulation details). For each $\delta$, we draw an initial ensemble of $20\,000$ random states ($T = \infty$), and numerically evolve each state in time according to Eq.~\eqref{eq:H_gamma}, storing a snapshot every $\Delta t = J^{-1}$. We then calculate the correlator for each state directly, and finally average over the different states. We extract $\mD$ and $\lambda$ by fitting the spin autocorrelator \eqref{eq:spin_autocorr} with the crossover form and find the predicted $1/\delta$-scaling near the integrable limit, shown in Fig.~\ref{fig:gamma_superdiffusion}(a). We show that the spin autocorrelators are well-described by this crossover in Fig.~\ref{fig:gamma_superdiffusion}(b).

\sectionn{Superdiffusion timescales}
To quantify the crossover timescale, we  consider the effective dynamical exponent
\eqn{
\alpha_{\eff}(t) = -\frac{d\log \mA}{d\log t} = \frac{2}{3} - \frac{1/6}{1 + \lambda t^{-1/3}},
\label{eq:effective_exponent}
}
such that, locally in time, $\mA(t) \sim t^{-\alpha_{\eff}}$. We can define the crossover time $t^*$ as the time after which $\alpha_{\eff}(t) < \alpha^*$, for some (arbitrary) threshold $1/2 < \alpha^* < 2/3$. Regardless of the chosen threshold,  Eqs.~\eqref{eq:diffusion_coefficient_crossover} \& \eqref{eq:effective_exponent} posit
\eqn{
t^* \sim \lambda^3 \sim \delta^{-3}.
\label{eq:crossover_time_delta}
}
That is, superdiffusion becomes parametrically long-lived, with a sharp divergence in its lifetime, near the integrable point. Moreover, the above picture of a crossover  driven by decay in two-soliton scattering predicts the dependence on the degree of integrability-breaking $\delta$.

\sectionn{Multiple crossovers}
We note, however, that away from the integrable limit, say, for moderate $\delta \gtrsim 0.3$, the diffusion constant starts to noticeably deviate from the $\mD \sim 1/\delta$ prediction,  see Fig.~\ref{fig:gamma_superdiffusion} (inset).  Whilst a regime of logarithmically-enhanced diffusion cannot be well-separated from the crossover to ordinary diffusion,  processes that cause soliton decay will at any rate suppress $\mD$. 

We consider again the scattering of large solitons with a gas of smaller excitations, treating this as an effective noisy perturbation. The scaling of the decay rate with $s$ is set by the Goldstone bound \cite{de2021stability}, and the decay matrix elements should still scale with $\delta$. We therefore modify Eq.~\eqref{eq:crossover} with an additional decay factor $e^{-\gamma \delta^2 t/s^2}$, for some constant $\gamma$, which changes the diffusion constant to
\eqn{
\frac{1}{\mD(\delta)} \sim \frac{2\eta^{1/2}\mu^{1/2}}{\pi\kappa}\delta + \frac{2\gamma}{\pi^2\kappa}\delta^2. 
\label{eq:diffusion_constant_both_channels}
}
We find numerically that, for our parametrisation $\delta$, the linear and quadratic terms of Eq.~\eqref{eq:diffusion_constant_both_channels} appear with roughly equal coefficients (Fig.~\ref{fig:gamma_superdiffusion}~(inset)). 

We note that if, say, the two-soliton processes were suppressed for some reason, a crossover from superdiffusion to logarithmically-enhanced diffusion would be parametrically \textit{slower}, ${t^* \sim \delta^{-6}}$ \cite{mccarthy2024slow}. We expect that the $t^* \sim \delta^{-3}$ crossover to ordinary diffusion is the generic scenario, which prevents the logarithmically-enhanced diffusion from appearing as a well-separated temporal regime (cf. Fig.~\ref{fig:gamma_superdiffusion}), but for particular systems the logarithmically-enhanced regime may be especially prominent \cite{mccarthy2024slow}.


\begin{figure}
    \centering
    \includegraphics{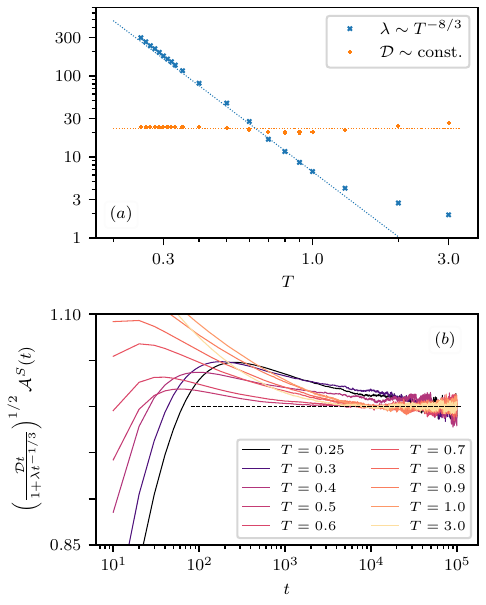}
    \caption{Long-lived superdiffusion at low temperatures in the (non-integrable) classical Heisenberg chain \eqref{eq:H_Heisenberg}. (a) The parameters $\lambda \sim T^{-8/3}$ and $\mathcal{D} \sim \mathrm{const}$, obtained by fitting the spin autocorrelations to the crossover Eq.~\eqref{eq:diffusion_coefficient_crossover}. (b) The spin autocorrelators at various temperatures, rescaled according to the crossover form \eqref{eq:diffusion_coefficient_crossover}, showing that they are consistent with this form -- though, at low temperatures, a ballistic regime (not accounted for by Eq.~\eqref{eq:diffusion_coefficient_crossover}) means that the crossover form is only attained at later times. 
    }
    \label{fig:Heisenberg_superdiffusion}
\end{figure}

\sectionn{Superdiffusion at finite temperature}
We turn now to the temperature-dependence of the superdiffusive timescales, focusing on the classical Heisenberg chain,
\eqn{
\mH = -J \sum_i \left( \bS_i \cdot \bS_{i+1} - 1 \right).
\label{eq:H_Heisenberg}
}
Despite being non-integrable -- and, arguably, the simplest interacting model of magnetism -- it evinces a long-lived regime of KPZ-like superdiffusion at finite temperature~\cite{mcroberts2022anomalous}. 

Again, we average over $20\,000$ initial states for each temperature. We expect that the main process driving the crossover to ordinary diffusion will, again, be decay in two-soliton collisions, so we fit the autocorrelation function at each temperature to the crossover form given by Eq.~\eqref{eq:diffusion_coefficient_crossover}, and extract the temperature dependence of the parameters $\mD(T)$ and $\lambda(T)$. 

We show this temperature scaling in Fig.~\ref{fig:Heisenberg_superdiffusion}(a). The lowest temperature for which these parameters can be directly extracted is $T \approx 0.25$; below this temperature, the KPZ-like superdiffusion persists for the entire simulation time window (up to $t = 10^5 J^{-1}$) \cite{mcroberts2022anomalous}. 

Further, for $T \lesssim 0.35$, the parameters estimated from the two-parameter fit of the numerical autocorrelator to Eq.~\eqref{eq:diffusion_coefficient_crossover} fluctuate significantly as a function of $T$, whereas we expect that they should be smooth. For $T \leq 0.35$ in Fig.~\ref{fig:Heisenberg_superdiffusion}, therefore, we fix $\mD \approx 23.5$ (the value extracted from the two-parameter fit at $T = 0.4$) and extract only an estimate of $\lambda(T)$ from the fit. 
In Fig.~\ref{fig:Heisenberg_superdiffusion}(b), we show the autocorrelators rescaled by the crossover with these parameters, which verifies that the autocorrelation data is consistent with fixing $\mD$ in this way.

In this framework, we find that the crossover from KPZ-like superdiffusion to ordinary diffusion in the classical Heisenberg chain is consistent with Eq.~\eqref{eq:diffusion_coefficient_crossover} and the parameter scalings 
\eqn{
\mD(T) \sim \mathrm{const.}, \;\;\; \lambda(T) \sim T^{-8/3},
}
and this scaling appears to hold up to temperatures ${T \approx J}$ (Fig.~\ref{fig:Heisenberg_superdiffusion}). In particular, the extracted scaling implies that the crossover timescale depends on temperature as
\eqn{
t^* \sim \lambda^3 \sim T^{-8}.
}

However, the low-$T$ dynamics is, in fact, richer than the crossover form Eq.~\eqref{eq:diffusion_coefficient_crossover}, which accounts only for the KPZ-like superdiffusion and the asymptotic ordinary diffusion. 

First, a ballistic regime becomes increasingly long-lived (and thus decreases the window in which Eq.~\eqref{eq:diffusion_coefficient_crossover} is a good description) as $T$ is lowered: the density of solitons decreases, which implies an increase in their screening lifetime, Eq.~\eqref{eq:KPZ_current}, up to which timescale each soliton contributes ballistically to spin transport. 

Second, besides the KPZ-like and diffusive regimes, the availability of the decay channel involving solitons of different sizes favours a regime of logarithmically-enhanced diffusion between these two. There is, at present, no theory providing sufficient quantitative guidance to allow fits distinguishing between these different scenarios, and, in any case, one would require well-separated timescales and hence a huge simulation/experimental time window, to properly resolve these and their crossovers. Indeed, we estimate that at $T = \infty$ in the Heisenberg chain, both decay channels make comparable contributions to the diffusion constant, Eq.~\eqref{eq:diffusion_constant_both_channels}, so that this competition will likely persist to finite $T$.

Moreover, since Eq.~\eqref{eq:crossover} does not seem to come with any obvious way to predict the temperature dependence of the coefficients, different scaling forms of the low-$T$ regime(s) are therefore not ruled out by our above analysis. In any event, however, we find that the timescale over which superdiffusion is observable has a very fast divergence at low temperature. If anything, attempting a fit of $\mD$ to the low-temperature data (not shown) seems to indicate that it grows upon lowering $T$, which would then lead an even more rapid divergence of $t^*$. All of this goes some way to explaining why there is no measureable deviation from the KPZ-like superdiffusion for temperatures ${T \lesssim 0.25J}$.

\sectionn{Conclusions}
Whilst ordinary diffusion is expected in the truly asymptotic limit for any non-integrable spin chain, the timescales associated with superdiffusive spin dynamics can be parametrically large -- and, as a question of experimental or numerical relevance, may push the crossover to diffusion to timescales at which it is effectively inaccessible, making KPZ-like superdiffusion not only easy to observe but in fact hard to avoid. In particular, the superdiffusive lifetime diverges very rapidly as a function of the degree of integrability-breaking (${t^* \sim \delta^{-3}}$), and particularly fast -- consistent with ${t^* \sim T^{-8}}$ -- at low temperature. 

We have related these timescales to the degrees of freedom responsible for the superdiffusion at integrability -- the large solitons -- and presented a possible mechanism of how the crossover to diffusion is driven by the decay of these solitons in collisions with each other. Clearly, much further work here is warranted, which may well fill in a picture of a rich landscape of non-universal crossovers. 

Indeed, integrability is an important concept across many classes of dynamical systems, but can never be perfectly realised in practice, and an understanding of how approximate integrability manifests itself is of clear practical and experimental relevance.
Whether the anomalous hydrodynamics associated with integrability can be similarly parametrically long-lived in different physical systems is a particularly intriguing question.

We re-iterate that the Heisenberg chain we have considered has not been especially chosen for the purpose of obtaining long-lived superdiffusive regimes; that this happens in, arguably, one of the simplest many-body models makes it all the more remarkable that anomalous hydrodynamics can persist over such enormous timescales. Indeed, the general question of just how common `non-generic' regimes in many-body systems are in practice is one of tremendous importance for experimental and numerical studies, and one which will keep us occupied for the foreseeable future.

This work was in part supported by the Deutsche Forschungsgemeinschaft under grants SFB 1143 (project-id 247310070) and the cluster of excellence ct.qmat (EXC 2147, project-id 390858490). 

{\bf Note added:} In the late stages of this project, we became aware of related work \cite{mccarthy2024slow} by Catherine McCarthy, Sarang Gopalakrishnan, and Romain Vasseur, who we thank for insightful discussions, particularly in pointing out the possible importance of the decay channel involving solitons propagating through an effective noisy background. They analyse the lifetime of the KPZ-like regime for periodic and noisy driving, while we consider Hamiltonian dynamics and its temperature dependence.



\bibliography{refs} 
\end{document}